\documentstyle[twocolumn,epsf]{jpsj}
\def \virg{\;\;,}
\def \point{\;\,.}
\def \kf{k_{\rm F}}

\def \i{{\rm i }}

\def \td{t_{\rm d}}

\def \W0c{W_{0 {\rm c}}}
\def \W0B{W_{0 {\rm B}}}
\def\ggs{\buildrel\textstyle > \over {\hbox{\raise0.2ex\hbox{$\sim$}}}}
\def\lls{\buildrel\textstyle < \over {\hbox{\raise0.2ex\hbox{$\sim$}}}}
\def\gsim{\,\lower0.75ex\hbox{$\ggs$}\,}
\def\lsim{\,\lower0.75ex\hbox{$\lls$}\,}
\def \sgn{{\rm sgn}}
\def\runtitle{
 Role of  Phase Variables in   Quarter-Filled 
 Spin Density Wave States
  }
\def\runauthor
{
 Yuh {\sc Tomio} and 
 Yoshikazu {\sc Suzumura} 
}

\renewcommand{\theequation}{\arabic{section}.\arabic{equation}}

\title{   
 Role of  Phase Variables in   Quarter-Filled 
 Spin Density Wave States\footnote{to be published in J. Phys. Soc. Jpn. {\bf 69} No.3 (2000) 796}
  }

\author{
Yuh {\sc Tomio}$^1$,
   and Yoshikazu  {\sc Suzumura}$^{1,2}$
}

\inst{
$^{1}$Department of Physics, Nagoya University, Nagoya 464-8602 \\ 
$^{2}$CREST, Japan Science and Technology Corporation (JST) \\
}

\recdate{November 1, 1999}

\abst
{
 Several kinds of spin density wave (SDW) states with both 
  quarter-filled band and  dimerization   are  reexamined 
 for a one-dimensional system with 
 on-site, nearest-neighbor  and next-nearest-neighbor 
 repulsive interactions,   
  which has been investigated by Kobayashi {\it et al.}
  (J. Phys. Soc. Jpn {\bf 67} (1998) 1098).
 Within the mean-field theory,
  the ground state and the response to the density variation 
   are  calculated in terms of  phase variables, $\theta$ and $\phi$, 
   where $\theta$  expresses  the charge fluctuation of SDW
 and  $\phi$  describes the relative motion 
 between  density wave with up spin and that with  down spin respectively.
 It is shown  that the exotic state of coexistence of 2$\kf$-SDW and 
 2$\kf$-charge density wave (CDW) is followed by  4$\kf$-SDW 
 but not by 4$\kf$-CDW where $\kf$ denotes a Fermi wave vector.   
 The  harmonic potential  with respect to 
 the variation of $\theta$ and/or $\phi$  
  disappears  
 for the interactions, which lead to the boundary 
  between   the pure 2$\kf$-SDW state and  
    the corresponding coexistent state.
}

\kword
{
spin density wave, quarter-filled band, charge density,
 phase variable,  long range Coulomb interaction
  }

\begin{document}
\sloppy
\maketitle
%-----------------------------------------------
\section{Introduction}
\setcounter{equation}{0}
 Spin density wave (SDW) state has been studied extensively  
 for  quasi-one-dimensional organic conductors, 
 (TMTSF)$_2$X and (TMTTF)$_2$X, which are known as Bechgaard salt.
\cite{Jerome,Yamaji}
 Since the energy band of the salt is anisotropic and quarter-filled, 
 the SDW state is determined essentially  
  by the nesting along the one-dimensional direction 
 where  the corresponding Fermi wave vector, $\kf$, is  
  $\kf = \pi/4$ in the unit of lattice constant. 
  The ground state of such a SDW, which  has been studied by 
 NMR experiment, \cite{Delrieu,Takahashi} exhibits the periodic array of 
 magnetic moment along the one-dimensional axis 
  where the periodicity is four times as large as the lattice constant 
 and the magnitude varies every two lattice sites. 
 The degree of  dimerization, which reduces the system 
  effectively to the half-filled one, has been estimated in these salt.
\cite{Mila} 
  Another noticeable feature of this SDW state is the charge fluctuation, 
   which is  caused by the deformation of the phase of the density wave, 
    as seen in  the nonlinear electric  conductivity.\cite{Gruner}

Recently the X-ray scattering investigation has shown that  
   2$\kf$-SDW coexists with  2$\kf$-charge density wave (CDW)
     in the SDW state of  (TMTSF)$_2$PF$_6$ 
 at temperatures just below  the onset temperature of 
the SDW ordering.\cite{Pouget}
 The 4$\kf$ satellite reflection is also observed 
   for both (TMTSF)$_2$PF$_6$ and (TMTTF)$_2$Br.
 Further the  coexistent state of 2$\kf$-SDW, 2$\kf$-CDW and 4$\kf$-CDW 
  has been  verified 
 in the  temperature range 4-12 K corresponding to SDW1 phase 
  while CDW is absent in SDW3 phases at lower temperatures.
\cite{Kagoshima}

 Theoretical study for the coexistence of SDW and CDW 
   has been explored  based on the mean-field theory.
  Seo and Fukuyama\cite{Seo} 
   have found the coexistence of 2$\kf$-SDW and 4$\kf$-CDW by examining 
    the extended Hubbard model, which takes  into account both   
     nearest-neighbor repulsive interaction and dimerization. 
  The SDW state, which is followed by 4$\kf$-CDW, 
   appears when  the magnitude of nearest-neighbor interaction 
     is larger than a critical value. 
  Based on this model, the collective mode 
describing the charge fluctuation   was calculated. 
  The excitation  spectrum, which has a  gap due to commensurability, 
    becomes gapless for the nearest-neighbor interaction  
     leading to  the onset of the  coexistence.\cite{Suzumura_jps97}
 The noticeable state of the  coexistence of 2$\kf$-SDW and 2$\kf$-CDW 
   has been demonstrated by Kobayashi {\it et al.} 
   who introduced  the next-nearest-neighbor interaction.\cite{Kobayashi}  
 They  examined the SDW state as the function of both
   the nearest-neighbor interaction ($V_a$) 
  and  the next-nearest-neighbor  interaction ($V_2$) 
     and found the phase diagram, in which 
     such a coexistent state is located in the region with 
    large $V_a$ and large $V_2$. 
  Thus the model with the long range Coulomb interaction 
  reveals the ground state  with  variety of   SDW states. 
 However 
 it is not yet  clear if  4$\kf$-SDW state and/or 4$\kf$-CDW state 
  can coexist  in addition to the  2$\kf$-CDW. 

 In the present paper,  such  a calculation is performed  
  for a model proposed by Kobayashi {\it et al.},\cite{Kobayashi} 
 within the  mean-field theory.  
 The ground state is  examined in terms of  two kinds of phase variables 
$\theta$ and $\phi$, 
 where $\theta$ is a phase of 2$\kf$-SDW and $\phi$ is 
 the difference between a phase of density wave with up spin and 
 that with down spin.  
  Further, as a first step for studying the density fluctuation,  
 we examine  the increase of  energy, which is caused 
 by varying  these two kinds of phases 
   from those of the ground state.
   In \S 2, formulation is given for the  model,
 which comprises   dimerization  and  
 alternating transfer energy ($t_a$ and $t_b$) with repulsive interaction 
($U$), the alternating nearest-neighbor interaction ($V_a$ and $V_b$)
 with the fixed $V_b/V_a$ 
 and the next-nearest-neighbor interaction ($V_2$).  
 The order parameters is examined  
  in terms of phase variable $\theta$ and $\phi$.  
 In \S 3, the ground state of the SDW is calculated to obtain 
  a phase diagram  on the plane of $V_2/V_a$ and  $V_a/t_b$. 
 Based on the study of the ground state,  
  the property of the density  variation  is investigated  
    by calculating  the energy as a function of $\theta$ and $\phi$. 
    Discussion is given in \S 4.

\section{Formulation}
 We study the SDW state with the quarter-filled band  
   by use of a  one-dimensional extended Hubbard model with 
    dimerization and  long range Coulomb interaction.
  The model is represented as 
%------------ (2.1)  ---------------
\begin{eqnarray}   \label{Hamiltonian} 
%	H  &=&  - \sum_{\sigma=\uparrow,\downarrow}  
%         \sum_{j=1}^{N}  
%         \left( t - (-1)^j \td  \right) 
%	 \left( C_{j\sigma}^\dagger  C_{j+1,\sigma} + h.c. \right) 
%	 + H_{\rm {int}}   \virg    \\
%        H_{\rm {int}}  &=&  U \sum_{j=1}^{N}
%               n_{j \uparrow} n_{j \downarrow}
%	     + \sum_{j=1}^{N} \left(  V - (-1)^j \delta V  \right)
%	       n_{j} n_{j+1}  
%	     + V_2 \sum_{j=1}^{N} n_{j} n_{j+2}  \virg
        H  &=&  - \sum_{\sigma=\uparrow,\downarrow}  
         \sum_{j=1}^{N}  
         \left( t - (-1)^j \td  \right) 
	 \left( C_{j\sigma}^\dagger  C_{j+1,\sigma} + {\rm h.c.} \right)
 \nonumber  \\
            & & 
	  {} + H_{\rm {int}}  \virg    \\  
        H_{\rm {int}}  &=&  U \sum_{j=1}^{N}
               n_{j \uparrow} n_{j \downarrow}
	     + \sum_{j=1}^{N} \left(  V - (-1)^j \delta V  \right)
	       n_{j} n_{j+1}   \nonumber \\
            & &
	  {} + V_2 \sum_{j=1}^{N} n_{j} n_{j+2}  \virg
\end{eqnarray}
where $t_a=t+\td$, $t_b=t-\td$, $V_a=V+\delta V$ and $V_b=V-\delta V$.
 The quantity  $C^\dagger_{j\sigma}$ ($C_{j\sigma}$) denotes 
  the creation (annihilation) operator of the electron 
    at the $j$-th lattice site with spin $\sigma$.  
     The quantity  $N$ is the total number of the lattice site. 
 $n_{j\sigma}=C^\dagger_{j\sigma}C_{j\sigma}$ and 
   $n_j=n_{j \uparrow}+n_{j \downarrow}$.  
We take $t$, $k_{\rm B}$ and the lattice constant as unity.  
 The alternation of the  transfer integrals given by  $t \pm \td$  
    comes from dimerization, which induces the   energy, $\td$. 
   Coupling constants for the Coulomb interaction are defined by 
      $U$, $V \pm \delta V$ and $V_2$, which  denote 
     the on-site repulsive interaction, the nearest-neighbor interaction 
      and the next-nearest-neighbor interaction, respectively. 
 The quantity $\delta V$ also originates in the dimerization.   
  By noting that the electron band is  at quarter-filling, 
    order parameters of density wave can be  written as
 ($ m=0,1,2,3$), 
%--------- (2.3a),(2.3b) ---------------  
{\setcounter{enumi}{\value{equation}}
\addtocounter{enumi}{1}
\setcounter{equation}{0} 
\renewcommand{\theequation}{\arabic{section}.\theenumi\alph{equation}}
\begin{eqnarray}   
   S_{mQ_0} &=& \frac{1}{N} \sum_{\sigma=\uparrow,\downarrow}  
                \sum_{-\pi < k \leq \pi } \sgn (\sigma)  
                \left<C_{k\sigma}^\dagger C_{k+mQ_0,\sigma}
                 \right>_{\rm MF}  \virg  \nonumber \\
		                  \label{OPS}   \\ 
   D_{mQ_0} &=& \frac{1}{N} \sum_{\sigma=\uparrow,\downarrow}  
                \sum_{-\pi < k \leq \pi }   
                \left<C_{k\sigma}^\dagger C_{k+mQ_0,\sigma}
                \right>_{\rm MF}     
                         \virg        \label{OPD}  
\end{eqnarray} 
\setcounter{equation}{\value{enumi}}}%
where
$Q_0 \equiv 2\kf=\pi/2$ with $\kf$ being the Fermi wave vector.
  The spatial dependence of electron densities with up and down spin,
 which  has the periodicity being  
  four times as large as the lattice constant,  is  obtained from 
   the Fourier transform of eqs. (\ref{OPS}) and (\ref{OPD}).
  Then we examine eqs. (\ref{OPS}) and (\ref{OPD}) under 
   the condition that  
 $S_0=0$, $D_0=1/2$, $S_{Q_0}=S^*_{3Q_0}$, $D_{Q_0}=D^*_{3Q_0}$, 
$S_{2Q_0}=S^*_{2Q_0} \equiv S_2$ and  $D_{2Q_0}=D^*_{2Q_0} \equiv D_2$.  
%--------------------------------------------------
 In terms of the  mean-field (MF) given  by  
   eqs. (\ref{OPS}) and (\ref{OPD}), 
Hamiltonian (eq. (\ref{Hamiltonian})) is expressed as  
%------------ (2.4) ------------
\begin{eqnarray}   \label{MFH} 
   H_{\rm MF} & = & 
	  \sum_{\sigma=\uparrow,\downarrow}   
         \sum_{-\pi< k \leq \pi }  
     \Biggl[ \left( \varepsilon_k + \frac{U}{4} + V + V_2 \right)
         C_{k\sigma}^\dagger C_{k\sigma}   
 \nonumber  \\ 
        & &  
\hspace{-7mm}
       + 
       \Biggl\{ \left( \bigl( \frac{U}{2} -2V_2 \bigr) D_{Q_0}
    + 2{\rm i} \delta V D^*_{Q_0} - \sgn(\sigma) 
       \frac{U}{2} S_{Q_0} \right) 
 \nonumber  \\
        & &
\hspace{2mm}
     {}  \times  
         C_{k+Q_0 ,\sigma}^\dagger  C_{k\sigma} + {\rm h.c.}  \Biggr\}
 \nonumber  \\
        & &
\hspace{-7mm}
       +
       \Bigl( \bigl( \frac{U}{2} -2V + 2V_2 \bigr) D_{2Q_0}
    - \sgn(\sigma) \frac{U}{2} S_{2Q_0} 
 \nonumber  \\
        & &
\hspace{2mm}
    {} - 2 {\rm i} \td \sin k \Bigr)
         C_{k\sigma}^{\dagger} C_{k+2Q_0,\sigma} \Biggr]   
 \nonumber  \\ 
        & &
\hspace{-7mm}
       + 
       NU \left[ - \frac{1}{16} -\frac{1}{2} \left( 
       {|D_{Q_0}|}^2 - {|S_{Q_0}|}^2 \right) \right.
 \nonumber  \\
        & &  \left.
\hspace{2mm}
     {} - \frac{1}{4} \left( D_{2Q_0}^{2} - S_{2Q_0}^{2} 
       \right) \right]
 \nonumber  \\ 
        & &
\hspace{-7mm}
       + 
       NV  \left(  -\frac{1}{4} + D^2_{2Q_0}  \right) 
     + {\rm i}N \delta V  \left(  D^2_{Q_0} - D^{*2}_{Q_0}  \right)
 \nonumber  \\
        & & 
\hspace{-7mm}
       + NV_2 \left( -\frac{1}{4} + 2 {|D_{Q_0}|}^2 
       - D^2_{2Q_0} \right)   \virg
%   H_{\rm MF} & = & 
%	  \sum_{\sigma=\uparrow,\downarrow}   
%         \sum_{-\pi< k \leq \pi }  
%     \Bigl[ \left( \varepsilon_k + \frac{U}{4} + V + V_2 \right)
%         C_{k\sigma}^\dagger C_{k\sigma}   
% \nonumber  \\
%        & &  
%       + 
%       \left\{ \left( \bigl( \frac{U}{2} -2V_2 \bigr) D_{Q_0}
%    + 2{\rm i} \delta V D^*_{Q_0} - \sgn(\sigma) 
%       \frac{U}{2} S_{Q_0} \right)    
%         C_{k+Q_0 ,\sigma}^\dagger  C_{k\sigma} + h.c.  \right\}
% \nonumber  \\
%        & &
%       +
%       \left( \bigl( \frac{U}{2} -2V + 2V_2 \bigr) D_{2Q_0}
%    - \sgn(\sigma) \frac{U}{2} S_{2Q_0} 
%    - 2 {\rm i} \td \sin k \right)
%         C_{k\sigma}^{\dagger} C_{k+2Q_0,\sigma} \Bigr]   
% \nonumber  \\ 
%        & &
%       + 
%       NU \left[ - \frac{1}{16} -\frac{1}{2} \left( 
%       {|D_{Q_0}|}^2 - {|S_{Q_0}|}^2 \right) 
%     - \frac{1}{4} \left( D_{2Q_0}^{2} - S_{2Q_0}^{2} 
%       \right) \right]
% \nonumber  \\ 
%        & &
%       + 
%       NV  \left(  -\frac{1}{4} + D^2_{2Q_0}  \right) 
%     + {\rm i}N \delta V  \left(  D^2_{Q_0} - D^{*2}_{Q_0}  \right)
% \nonumber  \\
%        & & 
%       + NV_2 \left( -\frac{1}{4} + 2 {|D_{Q_0}|}^2 
%       - D^2_{2Q_0} \right)   \virg
\end{eqnarray}%
  where  $\varepsilon_k=-2t \cos k$. 
%--------------------
Since the nesting vector is given by  $Q_0 = \pi/2$, 
 the eigenvalue of eq. (\ref{MFH}) exhibits four energy bands 
  $E_{\rm n}(k)$, $({\rm n}=1,2,3,4)$, 
   where $E_1(k) < E_2(k) < E_3(k) < E_4(k)$.
In the ground state, 
  the lowest energy band, $E_1(k)$, is fully occupied and 
   others are empty due to the quarter-filling. 
 By solving eqs. (\ref{OPS}) and (\ref{OPD}) self-consistently, 
   the ground state energy per site, $E_{\rm {g}}$, is obtained   as
%-------------- (2.5) -------------  
\begin{eqnarray}   \label{Eg}
 E_{\rm g}  &=&  \frac{1}{N} \sum_{\sigma}   
         \sum_{0< k \leq Q_0 } E_1(k)     \nonumber  \\
        & &
\hspace{-6.5mm}
       + 
       U \left[ - \frac{1}{16} -\frac{1}{2} \left( 
       {|D_{Q_0}|}^2 - {|S_{Q_0}|}^2 \right) 
     - \frac{1}{4} \left( D_{2Q_0}^{2} - S_{2Q_0}^{2} 
       \right) \right]
 \nonumber  \\ 
        & &
\hspace{-6.5mm}
       + 
       V  \left(  -\frac{1}{4} + D^2_{2Q_0}  \right) 
     + {\rm i} \delta V  \left(  D^2_{Q_0} - D^{*2}_{Q_0}  \right)
 \nonumber  \\
        & & 
\hspace{-6.5mm}
       + V_2 \left( -\frac{1}{4} + 2 {|D_{Q_0}|}^2 
       - D^2_{2Q_0} \right)   \virg
\end{eqnarray}
where the Brillouin zone is reduced to $0 <  k \leq Q_0$.
In terms of eq. (\ref{Eg}), we define  the quantity, $\delta E$, as  
%------------ (2.6) --------------  
\begin{eqnarray}   \label{deltaE} 
 \delta E  =  E_{\rm {g}}  -  E_{\rm {normal}}  \virg 
\end{eqnarray}
which denotes  the difference between 
 the ground state energy of  the ordered state
 and that of the normal state  i.e.,  $S_j = D_j = 0$ ( $j=1,2$).
In  eq. (\ref{deltaE}),
 $E_{\rm{normal}}$ 
( $ = -2\sqrt{2} t /\pi  + U/16 + V/4 + V_2/4$ ) 
 is the energy of the normal state. 
 
%----------------------
 Now we examine the  SDW state from the aspect of the phase variable. 
Since  $S_{Q_0}=S^*_{3Q_0}$ and  $D_{Q_0}=D^*_{3Q_0}$, 
 eqs. (\ref{OPS}) and (\ref{OPD}) are expressed as
%--------------(2.7a), (2.7b) ------------------  
{\setcounter{enumi}{\value{equation}}
\addtocounter{enumi}{1}
\setcounter{equation}{0} 
\renewcommand{\theequation}{\arabic{section}.\theenumi\alph{equation}}
\begin{eqnarray}
S_{Q_0}  &=&  S_1 {\rm e}^{{\rm i}\theta} \virg  
   \hspace{0.5cm}  \label{SQ1}  \\  
D_{Q_0}  &=&  D_1 {\rm e}^{{\rm i}\theta^\prime}  \virg                
                   \label{DQ1} 
\end{eqnarray} 
\setcounter{equation}{\value{enumi}}}%
where $S_1$( $>$ 0) and $D_1$( $>$ 0) are real. 
 Quantities $\theta$ and $\theta^\prime$ are phase factors. 
 Within  the present numerical calculation, we find always 
 $S_{Q_0}/S_1 = \i D_{Q_0}/D_1$  for  the ground state.
 Thus  we calculate the energy  under such a restriction, i.e.,  
    eq. (\ref{DQ1}) is replaced by 
%---------------  (2.8) ------------------
\begin{eqnarray}
D_{Q_0}  =  D_1 {\rm e}^{{\rm i} (\theta -\pi/2)}  \point                
                   \label{DQ1D}  
\end{eqnarray}    
In terms of eqs. (\ref{OPS}), (\ref{OPD}), 
(\ref{SQ1})  and (\ref{DQ1D}), 
 the electron density for spin $\sigma$  at the $j$-th site 
  is expressed as 
%---------------------- (2.9) ------------------
\begin{eqnarray}
\left<n_{j\sigma}\right> &=& 1/4 
        + \sgn(\sigma) S_1 \hspace{-0.5mm} 
  \cos(Q_0 r_j \hspace{-0.5mm} + \hspace{-0.5mm} \theta)  
        + D_1 \hspace{-0.5mm} 
  \sin(Q_0 r_j \hspace{-0.5mm} + \hspace{-0.5mm} \theta) 
 \nonumber  \\
 & & {} + \frac{1}{2} \left( D_2 + \sgn(\sigma) S_2 \right)
          \cos(2Q_0 r_j) \point   \label{Density}  
\end{eqnarray} 
 The respective electron  density  is rewritten as
%------------- (2.10a), (2.10b) --------------
{\setcounter{enumi}{\value{equation}}
\addtocounter{enumi}{1}
\setcounter{equation}{0} 
\renewcommand{\theequation}{\arabic{section}.\theenumi\alph{equation}}
\begin{eqnarray}
\left<n_{j\uparrow}\right> &=& 1/4 
                    + A \sin(Q_0 r_j + \theta + \phi )  
                    + B \cos(2Q_0 r_j) \virg  \nonumber \\
\label{DensityU}  \\
\left<n_{j\downarrow}\right> &=& 1/4 
                    + A \sin(Q_0 r_j + \theta - \phi )  
                    + C \cos(2Q_0 r_j) \virg  \nonumber \\
\label{DensityD}  
\end{eqnarray}
\setcounter{equation}{\value{enumi}}}%
where $A = \sqrt{S_1^2 + D_1^2}$, $\phi = \tan^{-1} (S_1/D_1)$
 , $B =(D_2+S_2)/2$  and $C= (D_2-S_2)/2$.  
Equations (\ref{DensityU}) and (\ref{DensityD}) imply that 
  the density wave with up (down) spin has a phase 
 $\theta + \phi$ $(\theta -\phi)$.
The quantity,   $\phi$,  expresses the relative motion 
 between the density wave with up spin 
 and that with the down spin.
 The idea, that $\phi \not= 0$ leads to 
  the coexistence of 2$\kf$-SDW and  2$\kf$-CDW,  
  has been already asserted  by Overhauser.
\cite{Overhauser}  
 The quantity,  $\theta$,   
 denotes a phase representing the charge  density fluctuation, which  
 gives rise to  the sliding motion of the SDW.
\cite{Lee,Takada,Maki,Suzumura_I,Suzumura2}

 The calculation of the energy  can be implemented 
  by  two kinds of methods,  i.e., 
   parameters determined self-consistently 
 are given  either by $S_1$, $S_2$, $D_1$, $D_2$ and $\theta$ 
 or by $A$, $S_2$, $D_2$, $\theta$ and  $\phi$. 
 For the simplicity,  the former choice is taken for the calculation 
 of the ground state  while the latter one is  
 used for studying the effect of 
 the variation of phases $\theta$ and $\phi$ 
  around the ground state. 
 We define $\tilde{V}$ as the energy  
 in the presence of  density waves with 
  $S_{nQ_0}$ and $D_{nQ_0}$ ($n$=1,2 and 3), 
 which are introduced as the trial function. 
  From eqs. (\ref{MFH}), (\ref{SQ1}) and  (\ref{DQ1D}) 
    together with  $\phi = \tan^{-1}(S_1/D_1)$ , 
      $\tilde{V}$ is written as   
%--------------  (2.11) --------------
\begin{eqnarray}  \label{tildeV} 
\tilde{V} \equiv \frac{1}{N} \left< H_{\rm {MF}} \right> 
                - \frac{U}{16} - \frac{V}{4} - \frac{V_2}{4}  \point
\end{eqnarray} 
The average, $\left< ~~ \right>$,  is performed 
  by  eq. (\ref{MFH}), in which quantities   
    $S_{nQ_0}$ and $D_{nQ_0}$ are retained as parameters.  
Note that  eqs. (\ref{OPS}) and (\ref{OPD}) are equivalent to
  the condition that 
%----------- (2.12), (2.13) ---------------------
\begin{eqnarray} 
 \frac{\partial \tilde{V}}{\partial S_j}  & = & 
 \frac{\partial \tilde{V}}{\partial D_j}  = 0  
   \virg  \hspace{1cm} ( j=1,2)  
   \virg    \label{dVdS}  \\ 
 \frac{\partial \tilde{V}}{\partial \theta}  
 & = & 0
   \virg    \label{dVdtheta}   
\end{eqnarray}
 and   
  $N^{-1} \sum_j \left< n_{j\sigma} \right> = 1/4$. 
 Then we introduce a quantity,  $V( \theta, \phi )$,  
 which is defined by 
%-------- (2.14) ----------------- 
\begin{eqnarray}  \label{Vthph} 
 V( \theta, \phi )  \equiv  \tilde{V} 
  \big| _{ 
\frac{\partial \tilde{V}}{\partial A} =  
\frac{\partial \tilde{V}}{\partial S_2} =  
\frac{\partial \tilde{V}}{\partial D_{2}} =  0
} 
 \point 
\end{eqnarray}
The minimum of  $V(\theta,\phi)$  is given by 
 $\theta =\theta_0$ and $\phi =\phi_0$, i.e., 
   quantities $\theta_0$ and $\phi_0$ are phases   for 
   the ground state.   
 Thus, we can evaluate  
   the increase of energy by $V(\theta,\phi)$
    when the phase variables are varied  from  the ground state.
  Especially  the limiting  behavior for  small variation of
  these phases  is evaluated  by expanding $V(\theta,\phi)$ in terms of 
   $ \theta - \theta_0 (\equiv r \cos \psi) $ and  
 $\phi - \phi_0 (\equiv r \sin \psi)$. 
 For  $r \ll 1$, one obtains  
%------------ (2.15) ---------------  
\begin{eqnarray}   \label{defCpsi}
 V ( \theta, \phi ) =  V ( \theta_0, \phi_0  ) 
                    + C(\psi) r^2  +  \cdots  \virg  
\end{eqnarray}
 where $C(\psi)$ denotes  the coefficient of harmonic potential.

%-------------------------------------------------
\section{SDW States and Phase Variables}
\setcounter{equation}{0}
\subsection{Ground state }

%---------------------  Fig. 1 --------------
 
  We examine  the possible ground state, which gives  
  the lowest energy in the solution of  the self-consistency equations. 
  From  the numerical calculation, we found    
     three kinds of solutions for SDW states shown in Fig. 1: 
   pure 2-$\kf$-SDW (I), 
   a coexistent state of 2$\kf$-SDW and 4$\kf$-CDW (II) and 
     that of  2$\kf$-SDW, 2$\kf$-CDW and 4$\kf$-SDW (III). 
   The  pure 2-$\kf$-SDW state 
  is obtained   for the interaction of only  $U$.
 \cite{Suzumura2}
 A  coexistent state   of 2$\kf$-SDW and 4$\kf$-CDW 
\cite{Seo}   appears  for   $V$ lager than a  critical value. 
 An exotic state of the  coexistence of 2$\kf$-SDW and  2$\kf$-CDW
   has been obtained for  large $V$ and large $V_2$.\cite{Kobayashi} 
 This  state   differs from the state III in the sense that  
   4$\kf$-SDW also coexists in the state III. 
  It seems to be reasonable that 
    the  state III could be possible 
       if  the order parameter for 4$\kf$-SDW 
          is treated in a perturbational method . 
%
%============ Fig. 1 =============
\begin{figure}[b]
\begin{center}
\vspace*{6mm}
\leavevmode
\epsfysize=7.5cm
   \epsffile{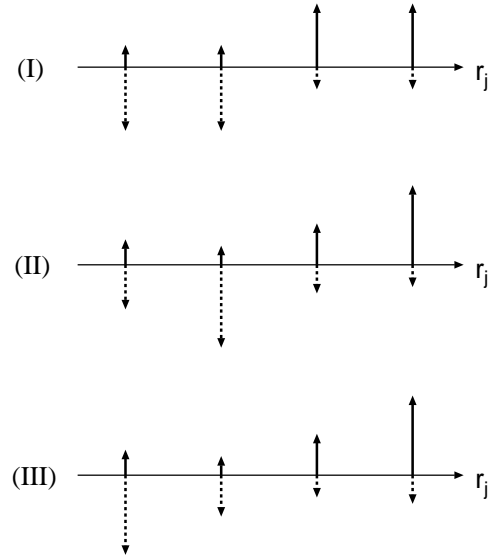}
\end{center}
%\epsfigure{file=f1.eps,height=7.5cm}
\vspace*{-2mm}
\caption{
 The electron density $<n_{j \sigma}>$ 
  as the  function of the lattice site $r_j$  
 for three  kinds of SDW states:  
  pure 2-$\kf$-SDW (I), 
   a coexistence of 2$\kf$-SDW and 4$\kf$-CDW (II) and 
     a coexistence of 2$\kf$-SDW, 2$\kf$-CDW and 4$\kf$-SDW (III). 
 The variation satisfies  $<n_{j+4}> = <n_{j}>$ 
  where the solid (dashed) arrow denotes the magnitude of 
  electron density with up (down) spin.
}
\end{figure}
%=================================
%

%------------- Fig. 2 ----------------
%
%============ Fig. 2 =============
\begin{figure}
\begin{center}
\vspace*{6mm}
\leavevmode
\epsfysize=7.5cm
   \epsffile{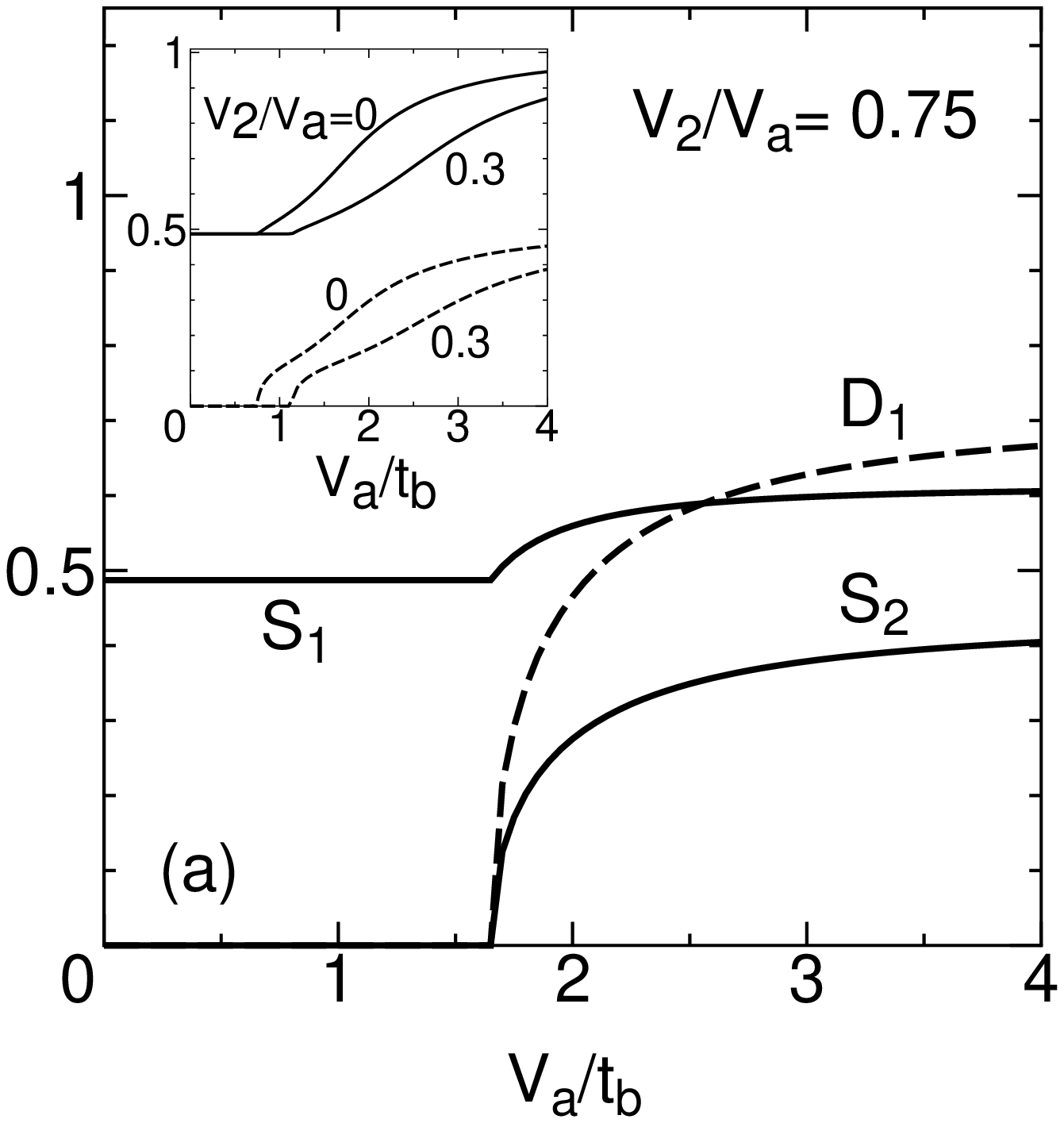}
\vspace*{6mm}
\epsfysize=7.5cm
   \epsffile{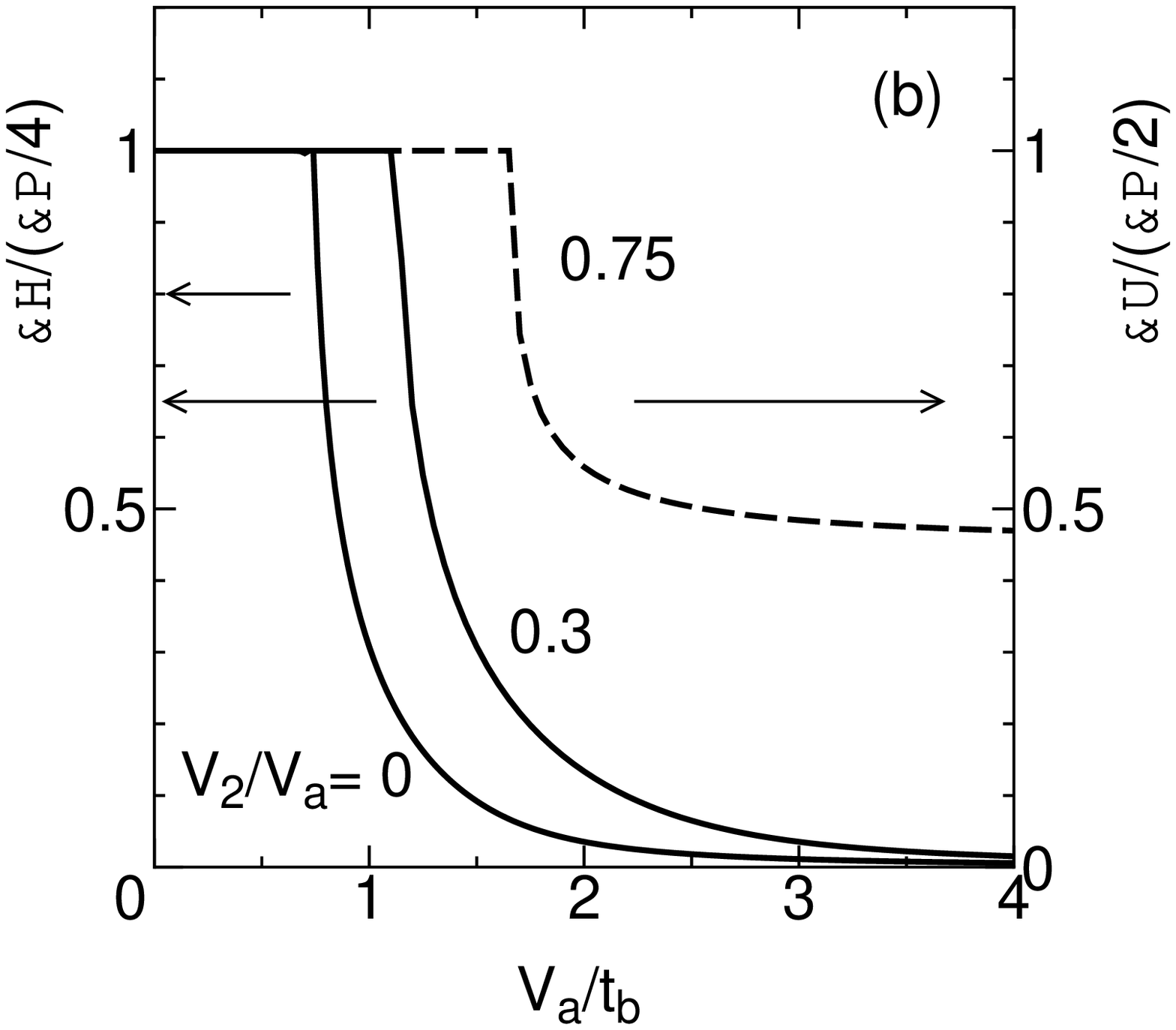}
\end{center}
%\epsfigure{file=f2a.eps,height=7.5cm}
%\epsfigure{file=f2b.eps,height=7.5cm}
\vspace*{-2mm}
\caption{
(a) The $V_a/t_b$-dependence of order parameters  
  $S_1$, $S_2$ and  $D_1$    
 for    $V_2/V_a = 0.75$.     
 In the inset, order parameters, 
  $S_1$ (solid curve) and $D_2$ (dashed curve) are shown for 
     $V_2/V_a$ = 0 and 0.3. 
 Order parameters are zero if not shown explicitly.          
 (b) The $V_a/t_b$-dependence of $\theta$ (solid curve) 
  for $V_2/t_b$ = 0 and 0.3 
  and that of $\phi$ (dashed curve) for   $V_2/t_b$ = 0.75.
}
\end{figure}
%=================================
% 
%
 In the present calculation,   
 we examine self-consistency equations 
     by varying  $V_a/t_b$ and $V_2/V_a$.  
  Parameters are chosen as $t_a/t_b=1.1$, $V_b/V_a=0.8$ and  $U/t_b =4$, 
\cite{Kobayashi} 
 where the magnitude of dimerization  seems 
    to  be reasonable for the Bechgaard salts.
\cite{Mila} 
 Figure 2(a) displays   the $V_a/t_b$-dependence of   
  $S_1$, $S_2$ and  $D_1$, 
    which  denote  amplitudes  for   2$\kf$-SDW,  4$\kf$-SDW and  
     2$\kf$-CDW  respectively .  
 For    $V_2/V_a = 0.75$,
  the state I moves to the state  III at $V_a/t_b = 1.66$, 
       where both $D_1$ and  $S_2$ starts increasing continuously 
 with a slight enhancement of $S_1$. 
  Such a simultaneous appearance of  $D_1$ and $S_2$  
   originates in the fact that   the term proportional to 
   $D_1 S_2 S_1$ exists in the expansion of 
  the   Ginzburg-Landau free energy. 
  Note that $D_2 = 0$, i.e., 
  4$\kf$-CDW  is  absent
  for  $V_2/V_a=0.75$. 
 In the inset, 
 the $V_a/t_b$-dependence of order parameters for 
     $V_2/V_a$ = 0 and 0.3 is shown 
  where solid curve and dashed curve denote 
     $S_1$ and $D_2$ respectively. 
  The transition from  the state I into   the state II takes place   
  at $V_a/t_b$ = 0.76 and 1.13  for $V_2/V_a$ = 0 and  0.3 respectively 
      where  $S_2 = D_1=0$ for any $V_a/t_b$. 
 It is found 
 that  $S_1$ in the state I does not depend  on  both $V_a$ and  $V_2$ 
 and that the state II is followed by 4$\kf$-CDW. 
% ---------  Fig. 2(b) --------------------
In Fig. 2(b), 
  quantity  $\theta$ corresponding to 
    $V_2/t_b$ = 0 and 0.3 is shown 
   by the solid curve. 
 When  the state I moves to the state II, 
   $\theta$  decreases rapidly from $\pi/4$ 
   and  $\phi$ remains at $\phi = \pi/2$ for both  states.  
The dashed curve denotes $\phi$ for    $V_2/t_b$ = 0.75, 
 which shows a transition from the state I into the state III. 
The quantity $\phi$ decreases from  $\pi/2$  
 while  $\theta = \pi/4$ for both states I and III.  
 Note that the coexistence of 2$\kf$-SDW and 2$\kf$-CDW 
 in the state III originates in $\phi$ with    $0 <\phi < \pi/2$.

%---------------   Fig. 3 ---------------------
%
%============ Fig. 3 =============
\begin{figure}[t]
\begin{center}
\vspace*{6mm}
\leavevmode
\epsfysize=7.5cm
   \epsffile{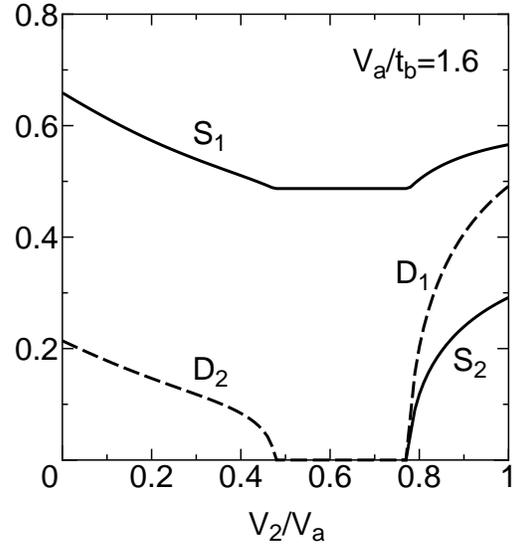}
\end{center}
%\epsfigure{file=f3.eps,height=7.5cm}
\vspace*{-2mm}
\caption{
The $V_2/V_a$-dependence of order parameters,
 $S_1$, $S_2$, $D_1$ and $D_2$,  for $V_a/t_b$ =1.6. 
}
\end{figure}
%=================================
%
%
  Figure  3 shows  
   the successive transition   from the state II, the state I and 
   the state III with increasing $V_2/V_a$.  
  Order parameters 
    $S_1$, $S_2$, $D_1$ and $D_2$ are shown as a function of 
       $V_2/V_a$ with the fixed $V_2/t_b$ =1.6. 
 A transition from the state II into state I occurs 
 at $V_2/V_a$ = 0.48  while  
 a transition from the state I into state III occurs 
 at $V_2/V_a$ = 0.78. 
 The $V_2$-dependence of $D_2$, $D_1$ and $S_2$  indicates  
    the transition being of the  second order.
 The quantities $D_1$ and $S_2$ appear 
  at  the same $V_2/V_a$, as seen in Fig. 2. 
   A linear dependence of the variation of $S_1$, which is found  
    just below  the transition at    $V_2/V_a$ = 0.48, 
       comes from the effect of higher order with respect to $D_2$. 
 In the state I,  $S_1$ does not depend on $V_2$ since  
  the pure 2$\kf$-SDW is determined only by the $U$-term, i.e.,   
   the density-density interaction of 
         both $V \pm \delta V$-term   and $V_2$-term   
      does  not give rise to  the  coupling 
        between the order parameter of   pure 2$\kf$-SDW  and others. 
 The appearance of $D_1$ and $D_2$ is understood as follows. 
  The distance  for the next-nearest-neighbor interaction, $V_2$,  
   corresponds  to  the 2$\kf$-periodicity. Then   
  the 2$\kf$-CDW is induced for $V_2$ 
 larger than a critical value, {\it e.g.},   
  $D_1$ appears for  $V_2/t \gsim 1.3 $
  and  irrespective of the magnitude of $V$.  
 In a similar way, one can understand 
 the role of  the nearest-neighbor interaction, $V$, which induces 
    $D_2$ for large $V_a$ (or $V$)
    due to  the 4$\kf$-periodicity. 
  The interval region in Fig. 3 corresponding to the state I  decreases 
  with increasing $V$ and disappears 
  for $V$ (and/or $V_a(=V + \delta V)$ ) 
     larger than a critical value.

%--------  Fig. 4 ------------------------
%
%============ Fig. 4 =============
\begin{figure}[t]
\begin{center}
\vspace*{6mm}
\leavevmode
\epsfysize=7.5cm
   \epsffile{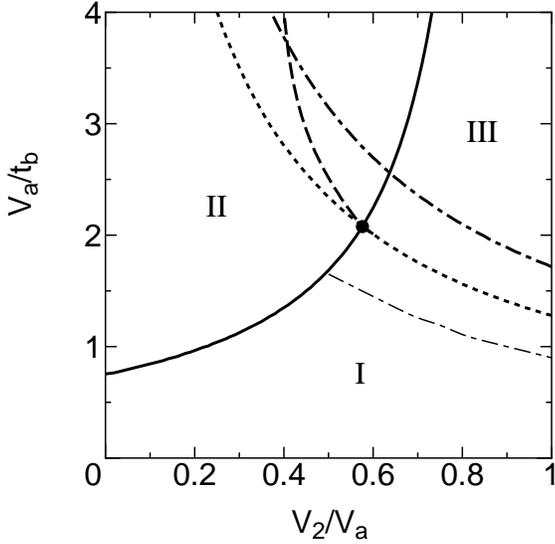}
\end{center}
%\epsfigure{file=f4.eps,height=8cm}
\vspace*{-2mm}
\caption{
 Phase diagram on the plane of 
     $V_2/V_a$  and $V_a/t_b$ 
     with $t_a/t_b = 1.1$, $V_b/V_a=0.8$ and $U/t_b = 4$.   
 The closed circle given by   
   $(V_2,V_a) = (V_{2,c},V_{a,c})$ 
   denotes an intersection of the solid curve and the dotted curve.  
 The solid curve (dotted curve), which  shows the boundary between 
 the region I and the region II (the region I and the region III) 
 becomes the actual boundary for  $V_a < V_{a,c}$.    
The boundary between the state II and the state III is 
  given by the dashed curve. 
 The thick  dash-dotted curve corresponds to the 
 dotted curve   with $S_2 = 0$
\cite{Kobayashi} 
   and  the thin dash-dotted curve is explained 
     in the inset of Fig. 7(b). 
}
\end{figure}
%================================
%
%
In Fig. 4, a phase diagram of SDW states 
 is shown  on the plane of 
 $V_2/V_a$  and $V_a/t_b$ for  $U/t_b = 4$, which is equal to
  $U/t$= 3.81.  
 The closed  circle corresponding to the intersection of 
   the solid curve and the dotted curve  denotes a critical point given by 
   $(V_{2,c}/V_a,V_{a,c}/t_b) (\simeq (0.576,2.077)$,  
  at which all the energies of states I, II and III become  equal. 
 The solid curve (the dotted curve) shows  the boundary between 
 the sate I and the state II (the state I and the state III). 
 The dashed curve is the boundary between 
  the state II and the state III.
  On the dashed curve,  
  the energy of the state II becomes equal  to that of the state III,
    while  amplitudes of both order parameters are finite. 
   For  $V_a < V_{a,c}$,   the second order transition 
  occurs  on  both the solid curve and    the dotted curve 
  while  the  first order transition from  the state II into
  the state III is obtained on the dashed curve.    
  We note that 
  the state III with the assumption of  $S_2=0$ 
   has been already obtained \cite{Kobayashi} 
     where the resultant boundary between 
       the state I and the state III is shown by 
        the thick dash-dotted curve in Fig. 4. 
 Compared   with the present boundary of the dotted curve, 
     the region for the state III is suppressed 
   in the  absence of $S_2$.  
 The thin dash-dotted curve, which is  not a boundary, 
   is explained   in the discussion  of the inset of Fig. 7(b). 

%
%---------------  Fig. 5 -------------------
%
%============ Fig. 5 =============
\begin{figure}[t]
\begin{center}
\vspace*{6mm}
\leavevmode
\epsfysize=7.5cm
   \epsffile{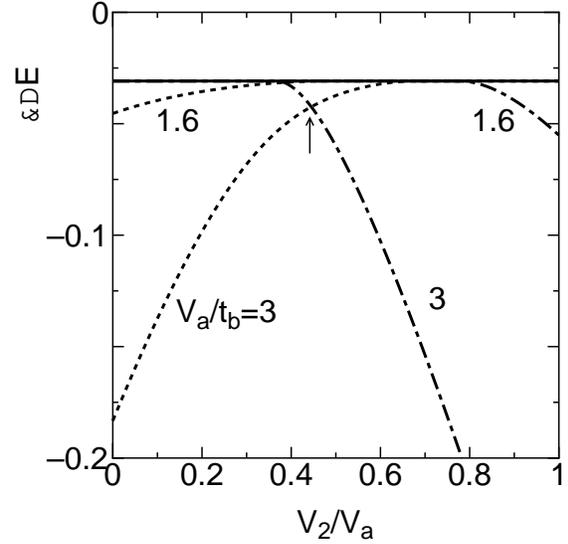}
\end{center}
%\epsfigure{file=f5.eps,height=7.5cm}
\vspace*{-2mm}
\caption{
The $V_2/V_a$-dependence of energy $\delta E$  ( eq. (\ref{deltaE})) 
  with the fixed  $V_a/t_b =$ 1.6 and 3,   
  for the state I (solid curve), 
 the state II (dotted curve) and the state III (dash-dotted curve).   
}
\end{figure}
%=================================
%
%
 The first order transition shown by  the dashed curve of Fig. 4 is 
 examined   by calculating the energy gain in the ordered state, 
  $\delta E$, which is defined by eq. (\ref{deltaE}).
 In Fig. 5,   the $V_2/V_a$-dependence of  $\delta E$ 
   is shown   for $V_a/t_b =$ 1.6 and 3.   
 The solid curve, the dotted curve and the dash-dotted curve   
  denote  $\delta E$ for 
 the state I, II and III respectively. 
 The quantity $\delta E$ for the state I, which  is independent of 
 the magnitude of $V_a/t_b$, is 
 largest compared with those of the state II and the state III.  
 The case of $V_a/t_b = 1.6$  corresponds to  Fig. 3,  
   which shows  a successive transition 
   given by   the state II, the state I 
 and the state III with increasing $V_2/V_a$.   
 For $V_a/t_b = 3$, 
 one obtains a first order transition from the state II into the state III
  at $V_2/V_a = 0.45$, which is shown by the arrow.  
 Based on these calculations,we find 
  the $V_2/V_a$ and $V_a/t_b$ dependences of   $\delta E$  in Fig. 4 
    as follows.
 The quantity $\delta E$ in the region I does not depend 
  on both $V_2/V_a$ and $V_a/t_b$ 
  while  $\delta E$ in region II (III) decreases motonically 
    with moving away from the boundary of the solid curve
     (the dotted curve).   

\subsection{Property around the ground state }

 In the previous section, the ground sate has been studied  
 by calculating order parameters where their  phases 
 $\theta(\rightarrow \theta_0)$ and $\phi(\rightarrow \phi_0)$ are also 
  determined so as to minimize the energy. 
 We examine, in this section,  the property  around the ground state 
  by calculating 
    $V(\theta,\phi)$ given by eq. (\ref{Vthph})
 which denotes the energy obtained  by  the phase variation.  
 The study is focused on the  state around  the boundary 
  given by  the solid curve and the dotted curve in Fig. 4.  

%---------------  Fig. 6 -------------------  
%
%============ Fig. 6 =============
\begin{figure}
\begin{center}
\vspace*{6mm}
\leavevmode
\epsfysize=7.5cm
   \epsffile{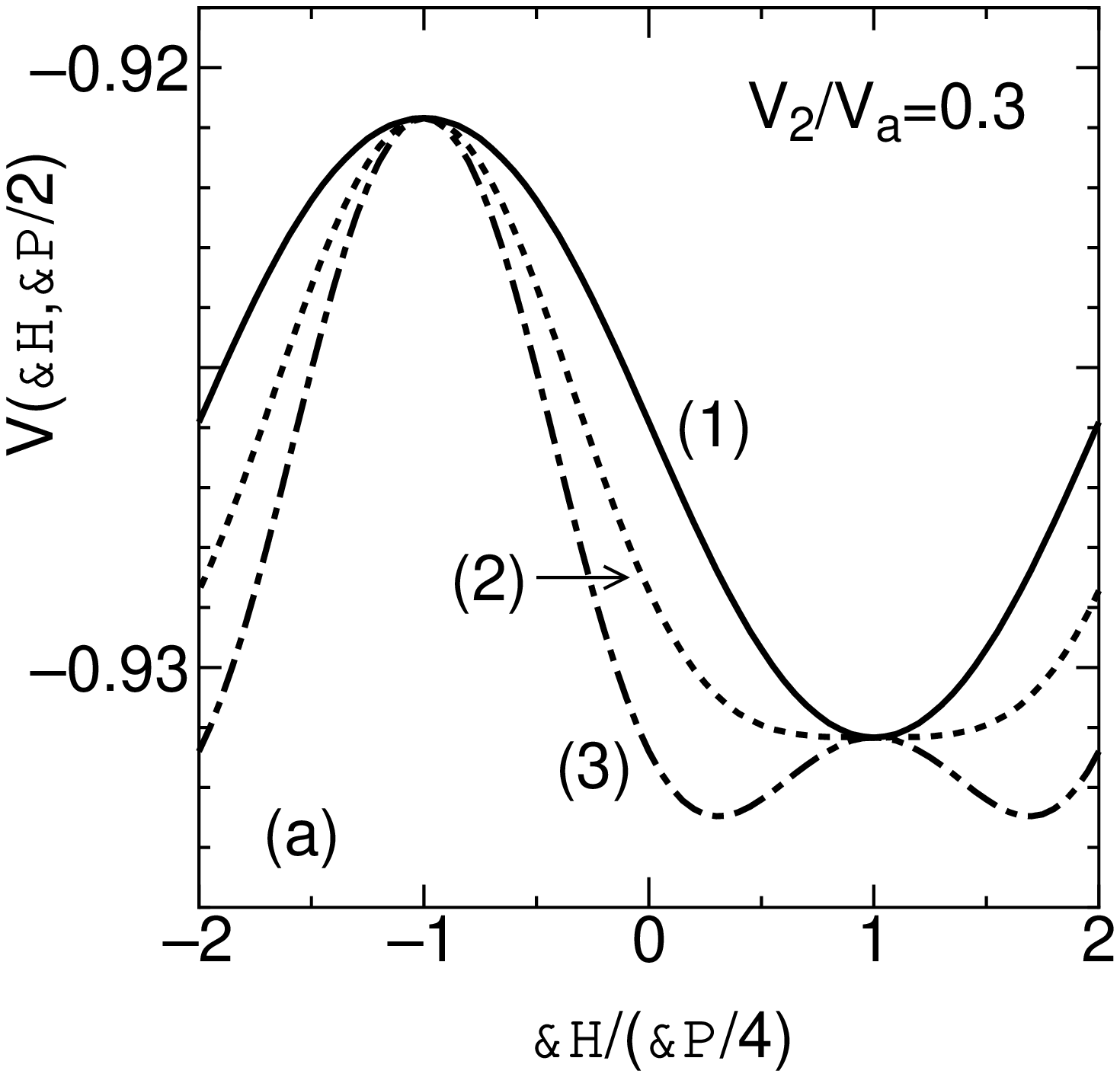}
\vspace*{6mm}
\epsfysize=7.5cm
   \epsffile{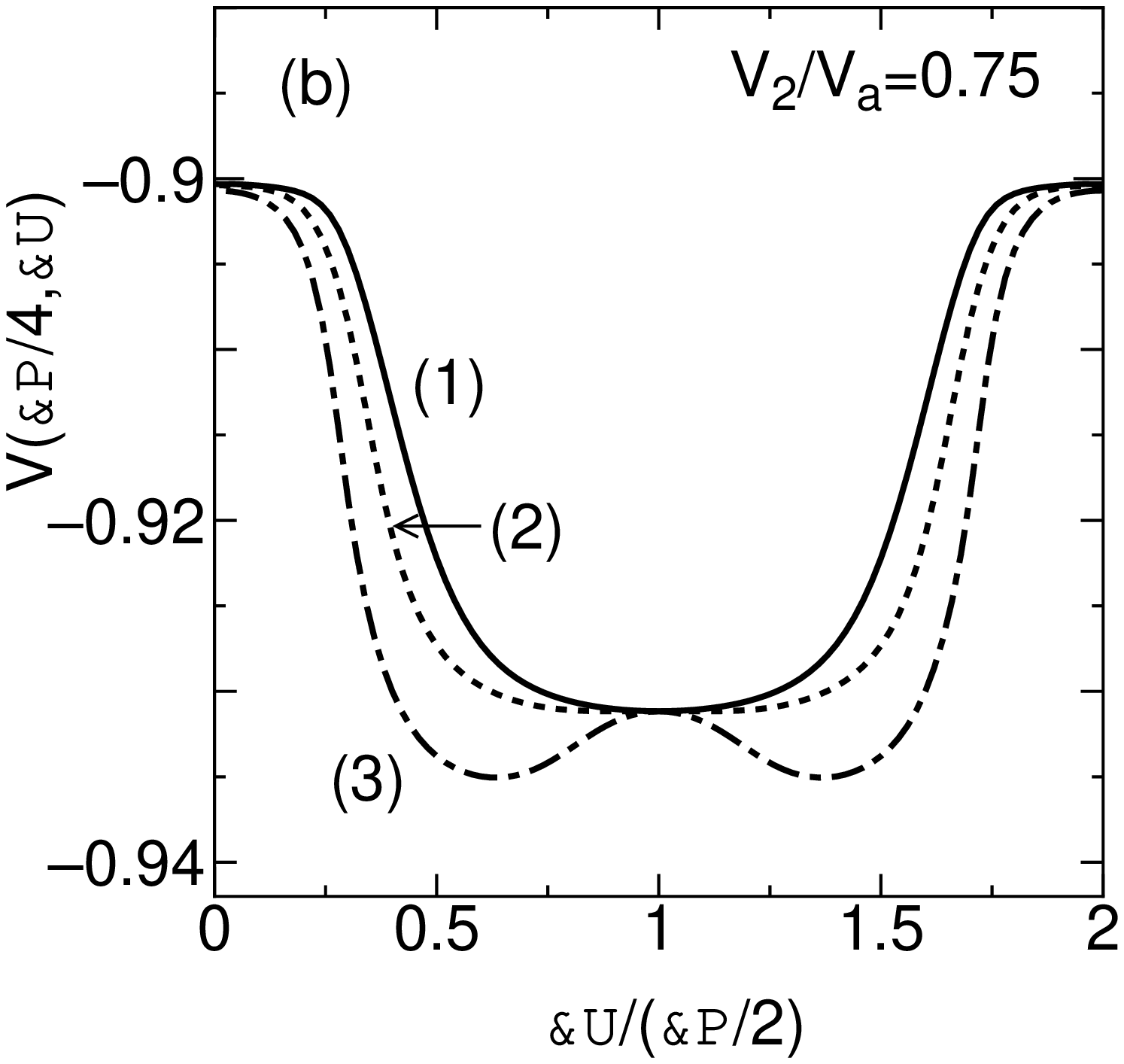}
\end{center}
%\epsfigure{file=f6a.eps,height=7.5cm}
%\epsfigure{file=f6b.eps,height=7.5cm}
\vspace*{-2mm}
\caption{
 (a) The $\theta$-dependence of $V(\theta,\pi/2)$ 
   for $V_a/t_b=$ 0.5 (1), 1.13 (2) and 1.5 (3) 
    with the fixed $V_2/V_a = 0.3$.   
 (b) The $\phi$-dependence of $V(\pi/4,\phi)$ 
      for $V_a/t_b=$ 1.4 (1), 1.66 (2) and 1.8 (3) 
        with the fixed $V_2/V_a = 0.75$. 
}
\end{figure}
%=================================
%
%
Figure 6(a) shows  the $\theta$-dependence of $V(\theta,\pi/2)$ 
   for three cases   
  which are  located close to the boundary of the solid curve of Fig. 4.  
 Solid curve, dotted curve and dash-dotted curve  are calculated  for  
   $V_a/t_b=$ 0.5 (1), 1.13 (2) and 1.5 (3) respectively 
       with the fixed  $V_2/V_a = 0.3$
   where   curve (2)  corresponds to the boundary 
     between the state I  and the state II. 
 The quantity $\partial^2 V(\theta,\pi/2) / \partial \theta^2$ 
    at $\theta = \theta_0$, 
 which denotes the second derivative of $V(\theta,\pi/2)$ with respect to 
   $\theta$  at the ground state,  is positive for  curves (1) and (3) 
     but   becomes zero  for  curve (2). 
 Such a result has been already found for   $V_2=0$ and $\delta V =0$.
\cite{Suzumura_jps97} 
 From Fig. 6(a), $V(\theta,\pi/2)$ can  be written as 
 $V(\theta,\pi/2 ) \simeq const. +  W_c \cos 4 \theta  + W_d\sin 2\theta$,
 where $W_c$ depends on both $V_a$ and $V_2$  and 
       $W_d$ is mainly  determined by  $\td$. 
 With increasing $V_a$ ($V_2$), $W_c$ decreases and becomes negative 
  (increases  and approaches to the limiting value with  $V_a=0$).
 In Fig. 6(b),  the $\phi$-dependence of $V(\pi/4,\phi)$ is shown for 
   the three cases  near the boundary of dotted curve of Fig. 4. 
  Solid curve, dotted curve and dash-dotted curve are calculated for 
    parameters  with      $(V_a/t_b,V_2/V_a)=$ (1.4,0.75) (1), 
     (1.66,0.75) (2) and (1.8,0.75) (3), which correspond to 
  the state I, the boundary and the state III  respectively.
For curve (2), the second derivative of $V(\pi/4,\phi)$ 
    with respect to   $\phi$ becomes zero at $\phi = \pi/2$ 
 while  harmonic potential exists  for curve (1) and (3). 
  Thus it turns out that  the harmonic potential  with respect to 
   $\phi$ vanishes only at   the boundary 
    between the state I  and the state III.

%---------------  Fig.7a  -------------------  
 
 It is of interest to examine  how the harmonic potential 
  for $\theta$ ($\phi$) reduces to zero  when parameters vary around 
     the boundary given by the solid curve (the dotted curve) in Fig. 4.
 We calculate  $C(\psi)$ given by eq. (\ref{defCpsi}), which denotes 
  a coefficient of harmonic potential with respect to 
    $\theta$ and $\phi$ around $\theta_0$ and $\phi_0$, i.e., 
      those of the ground state.  
 Quantities  $C(0)$ and $C(\pi/2)$ correspond to coefficients for 
    $\theta$   and $\phi$ respectively.
%
%============ Fig. 7 =============
\begin{figure}[t]
\begin{center}
\vspace*{6mm}
\leavevmode
\epsfysize=7.5cm
   \epsffile{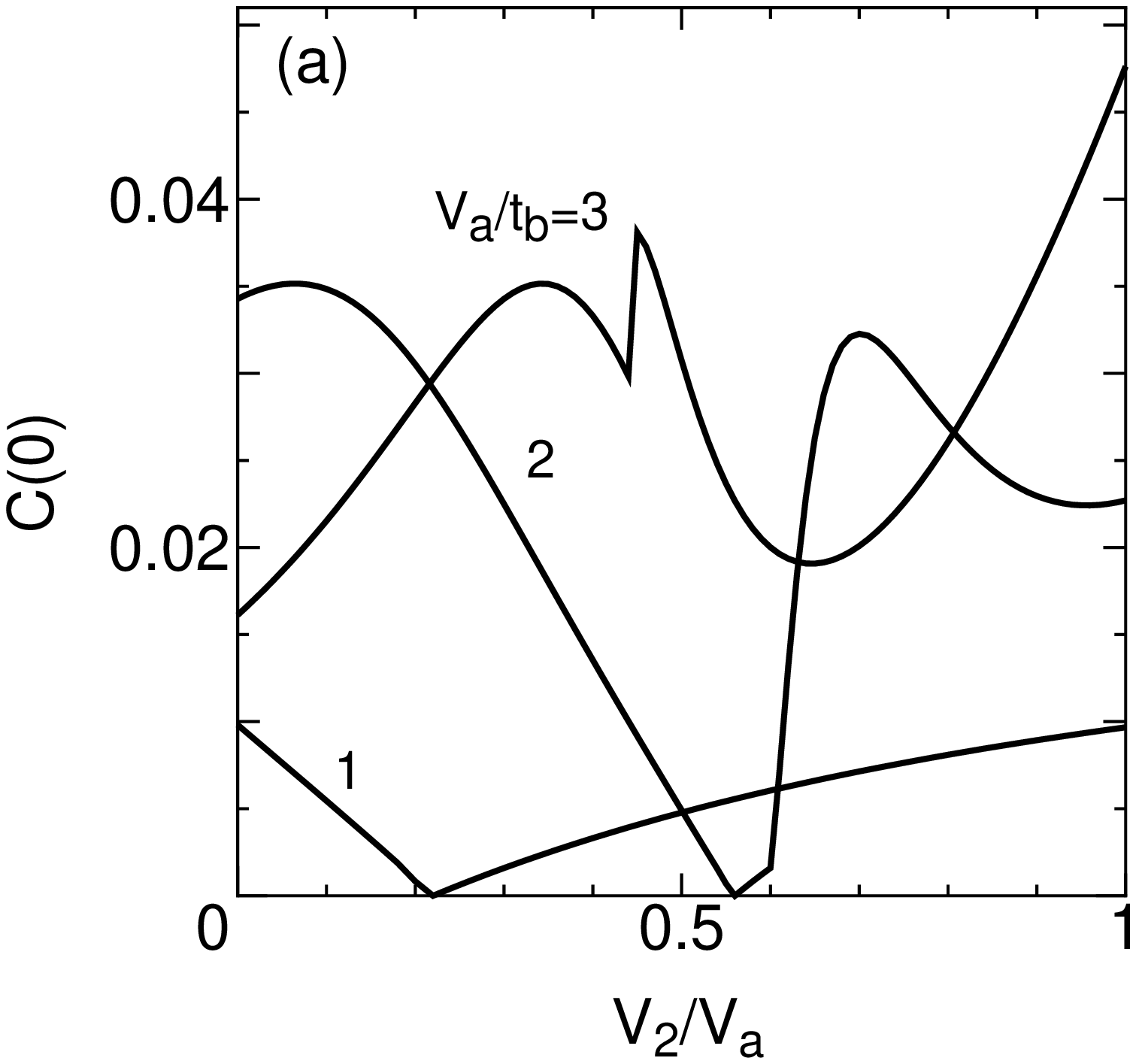}
\vspace*{6mm}
\epsfysize=7.5cm
   \epsffile{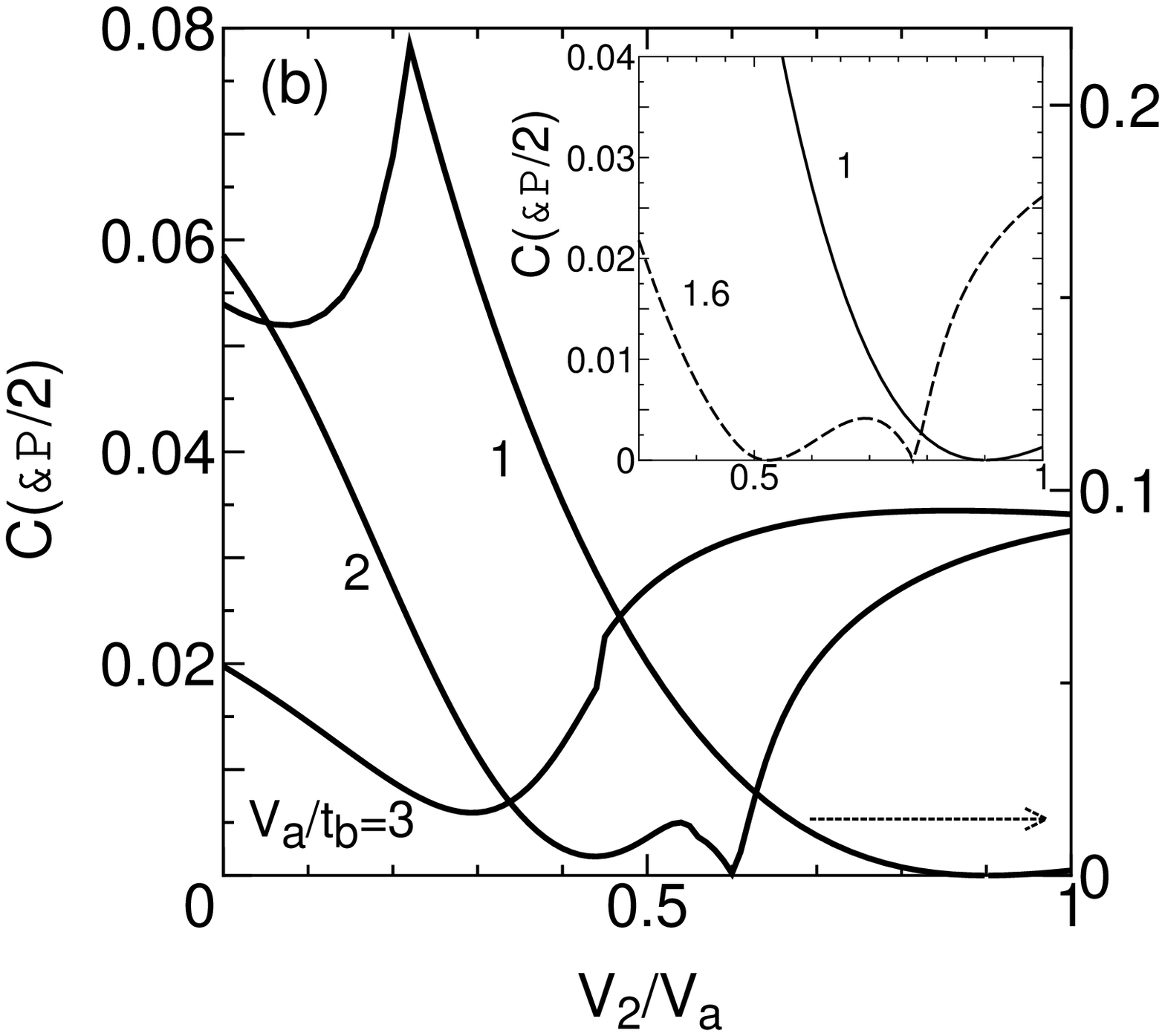}
\end{center}
%\epsfigure{file=fig7a.eps,height=7.5cm}
%\epsfigure{file=fig7b.eps,height=7.5cm}
\vspace*{-2mm}
\caption{
 (a)The $V_2/V_a$-dependence of $C(0)$ (eq. (\ref{defCpsi})) 
    for $V_a/t_b =$  1, 2  and 3.
%------------------  
(b)The $V_2/V_a$-dependence of $C(\pi/2)$ 
    for $V_a/t_b =$  1, 2 and 3 
      where the right vertical axis is for $V_a/t_b = 1$.       
 In the inset, the details of  $V_2/V_a$-dependence of $C(\pi/2)$ 
   is shown for $V_a/t_b$ = 1 and 1.6.  
}
\end{figure}
%=================================
%
 In Fig. 7(a),  the $V_2/V_a$-dependence of $C(0)$ is shown 
     for $V_a/t_b =$  1, 2 and 3 respectively.
 For $V_a/t_b =$ 1, $C(0)$ becomes zero at $V_2/V_a$ = 0.22, which 
    corresponds to  the boundary between the state I and the sate II. 
  Such a behavior  is similar to the $V$-dependence  of the charge gap, 
   which has been shown   for  $V_2 = 0$.\cite{Suzumura_jps97}   
 For $V_a/t_b =$ 2, $C(0)$ decreases to zero  at $V_2/V_a$ = 0.56, i.e.,  
    the boundary   between the state I and the state II
      and exhibits a   cusp at  $V_2/V_a$ = 0.61, i.e.,  
         the boundary between the state I and the state III.  
 A shoulder is found in  $C(0)$  for larger $V_2/V_a$. 
 For $V_a/t_b =$ 3, $C(0)$  exhibits  a jump, at which a first order 
  transition  takes place   due to  
  the boundary between the state II and the state III.  
 When $V_2$ increases further, 
    $C(0)$ takes a minimum and increases  rapidly.
%---------------  Fig.7b  ------------------- 
 In Fig. 7(b), the $V_2/V_a$-dependence of $C(\pi/2)$ 
   is shown for the same parameters as Fig. 7(a). 
 For $V_a/t_b =$ 1, $C(\pi/2)$ takes a cusp 
   at $V_2/V_a$ = 0.22 corresponding to a  transition 
    from the state II to  the state I. 
 With increasing $V_2/V_a$,  $C(\pi/2)$ decreases rapidly 
   and becomes  zero at $V_2/V_a = 0.90$ where 
      the quadratic behavior is seen around  zero.  
 The fact that such a zero value   does not imply 
   the boundary  is explained at the end of this paragraph. 
 For $V_a/t_b =$ 2, $C(\pi/2)$    shows a small cusp at 
     $V_2/V_a$ = 0.56 corresponding to a transition from 
     the state II to the state I. 
 A zero of  $C(\pi/2)$   at $V_2/V_a = 0.61$ comes from 
  the boundary between  the state I and the state III. 
 For $V_a/t_b =$ 3, $C(\pi/2)$ takes a  jump at  $V_2/V_a = 0.45$, 
  which originates in   the first order transition 
    between  the state II and the state III.    
%--------- inset ---------------
 In the inset, we show  the detail behavior around  the zero of 
  $C(\pi/2)$  for  $V_a/t_b$ = 1 and 1.6. 
 Quadratic behavior around the zero is seen for $V_a/t_b$=1. 
 There are two kinds of zeros for $V_a/t_b$ = 1.6 where   
 the larger $V_2$  corresponds  to  the transition from 
      the state I to the state III  and exhibits  
          a linear dependence    around  the zero.    
 For $V_a/t_b = 1.6$,  
  the  quadratic behavior is also found  around the smaller $V_2$, 
     at which the actual   transition does not occur. 
 Such a fact is understood as follows. 
 The effect of small variation of $\phi$ on 
  the expectation value of eq.(\ref{tildeV})
   is essentially determined by 
$
  \tilde{D}_{Q_0} \equiv 
   N^{-1} \sum_{\sigma=\uparrow,\downarrow}  
                \sum_{-\pi < k \leq \pi }   
                \left<C_{k\sigma}^\dagger C_{k+Q_0,\sigma}
                \right>  \virg 
$
 where the average $<>$ is performed over  eq.(\ref{MFH})   
   with $S_{mQ_0}$ and $D_{mQ_0}$ replaced by trial fields, {\it e.g.}, 
  $D_1 \propto \cos \phi$ and $S_1 \propto \sin \phi$.
  The ground state  of  the state I is the pure $2\kf$-SDW 
  with $\phi = \phi_0=(\pi/2)$  and then    
    quantities $D_{Q_0}$ and $D_{Q_0}^*$ in the second line 
      of eq. (\ref{MFH}) vanishes
  due to $D_1 \propto \cos \phi$ and eq. (\ref{DQ1}).  
  When $\phi$ is varied from $\phi_0=(\pi/2)$, 
   the magnitude of $D_{Q_0}$ and $D_{Q_0}^*$ becomes finite 
     and the resultant $\tilde{D}_{Q_0}$ 
 becomes finite  generally. 
 However there is a case where  $\tilde{D}_{Q_0} = 0$ even for 
   $D_{Q_0} \not= 0$   due to vanishing of the coefficient, {\it e.g.},
      $U/2-2V_2 = 0$ for $\delta V =0$.
 Such a case is examined explicitly  in the next section.  
 The location for  $C(\pi/2) =0$ 
 corresponding to  $\tilde{D}_{Q_0} = 0$
   is shown by the thin dash-dotted curve in Fig. 4,
   where the vanishing of $C(\pi/2)$ 
     exists only in the region I.
 The trajectory (thin dash-dotted curve) is nearly parallel to 
   the dotted curve and  crosses with the solid curve at 
       $(V_a/t_b, V_2/V_a) \simeq (1.6, 0.5)$. 

%---------------  Fig. 8 ----------------------
%
%============ Fig. 8 =============
\begin{figure}[t]
\begin{center}
\vspace*{6mm}
\leavevmode
\epsfysize=7.5cm
   \epsffile{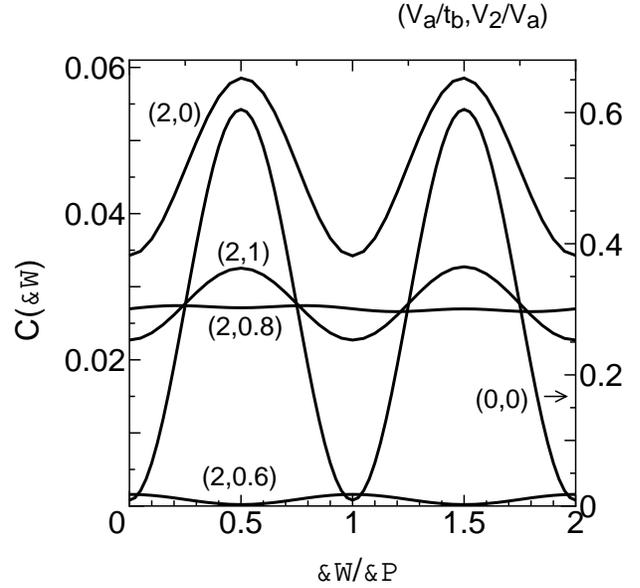}
\end{center}
%\epsfigure{file=fig8.eps,height=7.5cm}
\vspace*{-2mm}
\caption{
The  $\psi$-dependence of $C(\psi)$
 for $(V_a/t_b,V_2/V_a)$= (0,0), (2,0), (2,0.6), (2,0.8)
 and (2,1). 
 The right vertical axis is for (0,0).
}
\end{figure}
%=================================
%
%
 Further we examine $C(\psi)$ with arbitrary $\psi$, 
 which induces  the variation of 
     both $\theta$ and $\phi$.
  In Fig. 8, $C(\psi)$   is calculated 
     for $(V_a/t_b,V_2/V_a)$= (0,0), (2,0), (2,0.6), (2,0.8) and (2,1), 
      where  $C(\psi + \pi/2) = C(\psi)$. 
 The amplitude of  $C(\psi)$   is large   for $(0, 0)$ 
 (shown by  a vertical axis of right hand side ) but     
   becomes small in the presence of both  $V_a$ and $V_2$ 
    as seen from  (2, 0)  and (2,1). 
 The minimum of $C(\psi)$ is found at $\psi = 0$ for  (0,0), (2,0) and 
   (2,1) while it  is found  at  $\psi = \pi/2$ for (2,0.6). 
 Note that the minimum for  (2,0.8) is located between 
  $\psi=0$ and  $\psi=\pi/2$, where such a continuous variation of the 
    minimum  is seen   for  $0.80 < V_2/V_a < 0.86$ 
 and $V_a/t_b = 2$ in  the region III. 
  The minimum of $C(\psi)$ at $\psi =\pi/2$ is obtained for parameters 
  $(V_2/V_a,V_a/t_b)$ close to  the dotted curve in Fig. 4.     
  The quantity $C(\psi)$ exhibits a small $\psi$-dependence for 
     (2,0.8). 
  We comment on the behavior of $C(\psi)$ in Fig. 4. 
  Within the numerical accuracy of the present calculation, 
   we find $C(\psi) = 0$ at  a critical point 
     given by  $(V_a/t_b,V_2/V_a) = (2.077,0.576)$.  
 The small magnitude of $C(\psi)$  shown for  (2,0.6)  
   is  obtained for parameters of  $(V_a/t_b,V_2/V_a)$, which  
     are located near the critical point and also  
      in the region enclosed by  the dotted curve, solid curve 
        and thin dash-dotted curve.   
 Thus it is expected that  the fluctuations for $\theta$ and $\phi$ 
   become large around the solid curve and the dotted curve  
   respectively and that 
  the fluctuations for both $\theta$ and $\phi$ become large 
  around the critical point of the intersection of these two curves .

%------------------- Fig. 9 -------------------
%
%============ Fig. 9 =============
\begin{figure}[t]
\begin{center}
\vspace*{6mm}
\leavevmode
\epsfysize=7.2cm
   \epsffile{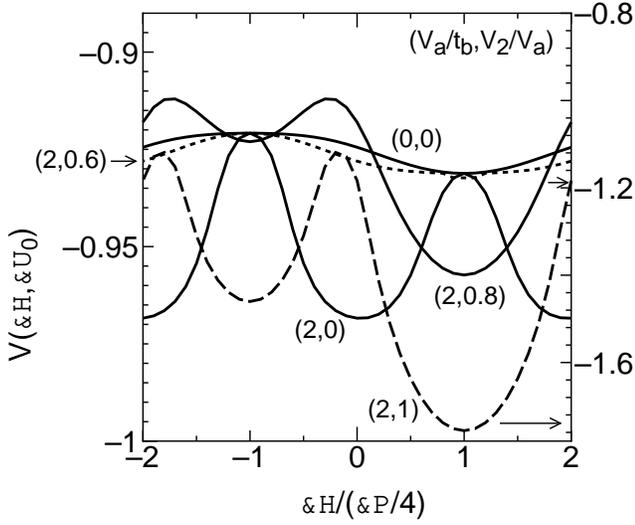}
\end{center}
%\epsfigure{file=fig9.eps,height=7.2cm}
\vspace*{-2mm}
\caption{
The $\theta$-dependence of $V(\theta,\phi_0)$ 
 for  $(V_a/t_b,V_2/V_a)$= (0,0), (2,0), (2,0.6), (2,0.8)
 and (2,1) where 
 $\phi_0 /\pi/2 = 1$ for   (0,0), (2,0) and, (2,0.6)  and 
 $ \phi_0/\pi/2 =$  0.537 and  0.499  for   (2,0.8) and (2,1) 
 respectively. 
}
\end{figure}
%=================================
%
%
 Now we examine  both  $\theta$ and $\phi$-dependences of $V(\theta,\phi)$ 
  around the ground state by choosing parameters  
   $(V_a/t_b,V_2/V_a)$= (0,0), (2,0), (2,0.6), (2,0.8)
 and (2,1), which are the same as Fig. 8.  
    In Fig. 9, the $\theta$-dependence of $V(\theta,\phi_0)$ 
 is shown with the fixed $\phi = \phi_0$ where   
   $\phi_0 /\pi/2 = 1$ for   (0,0), (2,0) and, (2,0.6)  and 
 $\phi_0/\pi/2 =$  0.537 and 0.499  for   (2,0.8) and (2,1) respectively. 
 There is a periodicity given 
    by  $V(\theta+\pi,\phi_0) = V(\theta,\phi_0)$.
 The $\theta$-dependence of $V(\theta,\phi_0)$ is small 
   for  (0,0) and (2,0.6) ( i.e.,   in the region I of Fig. 4 ) 
  while  it  is large for (2,0), (2,0.8) and (2,1)
 (i.e.,  in the region of II and III of Fig. 4). 
 The latter result  comes from the fact that  the variation of $\theta$ 
  needs a large amount of energy due to the presence of  
  the charge density of $D_2$  ($D_1$)  in  the region II (the region III).
%------------------  Fig. 10 -------------
%
%============ Fig. 10 =============
\begin{figure}[t]
\begin{center}
\vspace*{6mm}
\leavevmode
\epsfysize=7.5cm
   \epsffile{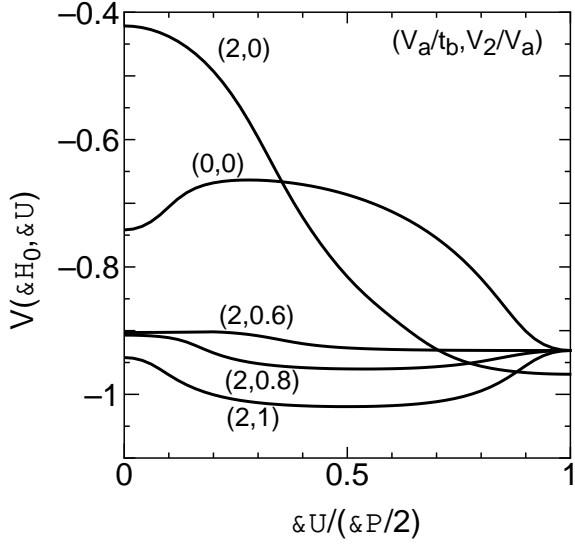}
\end{center}
%\epsfigure{file=fig10.eps,height=7.5cm}
\vspace*{-2mm}
\caption{
The $\phi$-dependence of $V(\theta_0,\phi)$ 
 for  $(V_a/t_b,V_2/V_a)$= (0,0), (2,0), (2,0.6), (2,0.8)
 and (2,1) 
 where  
 $\theta_0 /\pi/4 = 1$ for   (0,0), (2,0.6), (2,0.8) and (2,1)  and 
 $ \theta/\pi/4 =$  0.035  for   (2,0). 
}
\end{figure}
%=================================
%
 Figure 10 shows the corresponding  $\phi$-dependence  for 
  $V(\theta_0,\phi)$ where  
    $\theta_0 /(\pi/4) = 1$ for   (0,0), (2,0.6), (2,0.8) and (2,1)  and 
       $ \theta/(\pi/4) =$  0.035  for   (2,0)
where 
  $V(\theta_0,\phi) = V(\theta_0,\pi/2 - \phi) = V(\theta_0,\phi + \pi)$. 
 In the state III,   $\phi_0$ differs  from $\pi/2$, which causes a 
 small increase of  $V(\theta_0,\phi)$ for the variation of  $\phi$.     
 The variation of $V(\theta_0,\phi)$ is large for (2,0) and (0,0) 
    and is small for  (2,0.6), (2,0.8) and (2,1). 
 This indicates a fact that  the fluctuation of $\phi$  
  in the state I and II is small and that in the state III is large.
 The magnitudes of $V(\pi/4,\phi)$ for (0,0), (2,0.6), (2,0.8) and (2,1) 
   becomes equal at $\phi=\pi/2$,   
    since all  these states reduce to  the state I
    with the energy being independent of  both $V$ and $V_2$.

\section{ Discussion }
\setcounter{equation}{0}

In the present paper, we have examined several kinds of  SDW states  in 
 a one-dimensional quarter-filled electron system with long range Coulomb 
  interaction within the mean-field theory.
 The phase diagram of the  SDW state with the  states I, II and III 
  were  obtained on the plane of $V_2/V_a$ and $V_a/t_b$.
 The variation of the energy from that of the ground state was calculated 
 in order to examine the role of phase variables  of the SDW states.  

%-----------  Fig. 11 (a) -------------------------
%
%============ Fig. 11 =============
\begin{figure}[t]
\begin{center}
\vspace*{6mm}
\leavevmode
\epsfysize=7.5cm
   \epsffile{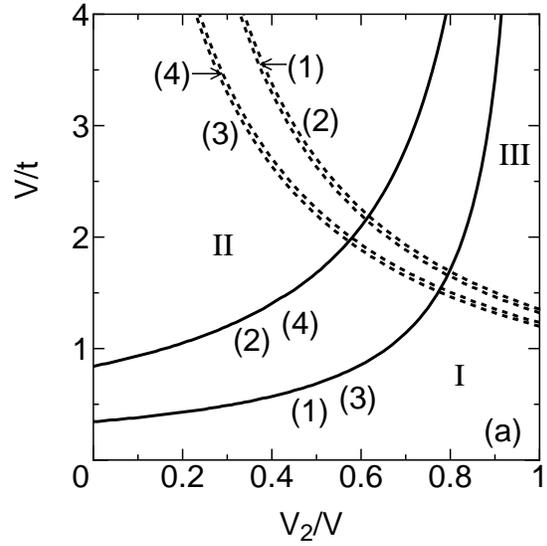}
\vspace*{6mm}
\epsfysize=7.5cm
   \epsffile{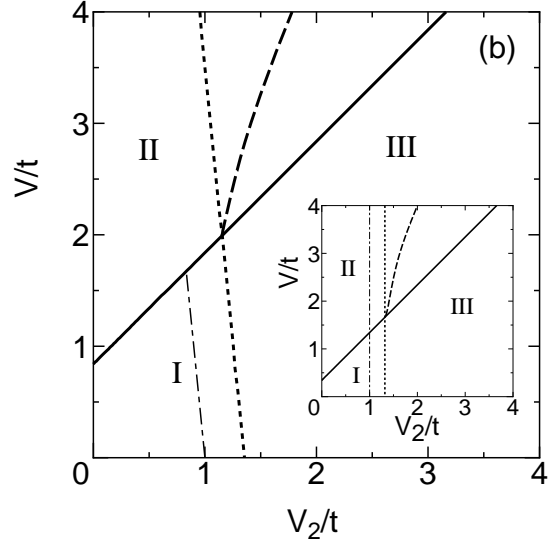}
\end{center}
%\epsfigure{file=f11a.eps,height=7.5cm}
%\epsfigure{file=f11b.eps,height=7.5cm}
\vspace*{-2mm}
\caption{
 (a)Phase diagram on the plane of $V_2/V$  and $V/t$ where 
   the solid curve ( the dotted curve ) denotes the boundary between  
    the state I and the state II ( the state I and the state III). 
 Parameters are taken as 
  $(\td/t, \delta V/V)$ = (0,0) (1), (0.1,0) (2), (0,0.1) (3), (0.1,0.1) (4).
(b)Phase diagram on the plane of $V_2/t$  and $V/t$ where 
   $U/t = 4$ and    $(\td/t, \delta V/V)$ = (0.1,0.1). 
 The notation of curves are the same as Fig. 4.  
 In the inset, another phase diagram  is shown  
  for $U/t = 4$ and    $(\td, \delta V)$ = (0,0).
}
\end{figure}
%=================================
%
 Here we examine the effect of $\delta V$ and $\td$ 
  on the boundaries in the phase diagram of Fig. 4. 
 In Fig. 11(a), the several results of 
   boundaries  are shown where 
   the solid curve ( the dotted curve ) denote the boundary between  
    the state I and the state II ( the state I and the state III). 
 Parameters are taken as  $(\td/t, \delta V/V)$ 
    = (0,0) (1), (0.1,0) (2), (0,0.1) (3), (0.1,0.1) (4) 
      where  curve (4) corresponds to Fig. 4.
 Since  solid  curve (2) ( solid curve (1)) is the the same as 
   solid curve (4) ( solid curve (3)),
   the boundary between the state I and the state II depends on 
    $\td$ but does not depend on $\delta V$. 
 The energy gain by the formation of 4$\kf$-CDW decreases by the 
 dimerization but not by the alternation of nearest-neighbor interaction.  
 From the dotted curves where curves (3) and (4) are located 
  below curves (1) and (2), it is found that 
  the energy gain by the formation  of 2$\kf$-CDW is 
   increased by $\delta V$. 
  The dotted curve moves slightly upward in the presence of $\td$. 
 We note that the comparison of 
  solid curve (4) and solid curve (3) 
   (dotted curve (4) and dotted curve (2))
    demonstrates  explicitly  the argument by
    Kobayashi et {\it al.}
\cite{Kobayashi} 
 that  the region II  is enlarged by in the absence of $\td$ 
    ( and that $\delta V$ increases the region III). 
%--------------   Fig. 11(b)-----------------

 In order to understand the coexistent  state clearly, 
  another phase diagram   is calculated on the plane of $V_2/t$ and $V/t$
 where $V=(V_a+V_b)/2$ and $t=(t_a+t_b)/2$. 
   by choosing  another parameters for both  axes. 
  In Fig. 11(b), a phase diagram corresponding to Fig. 4 
   is shown  with  $U/t = 4$ and  $(\td/t, \delta V/V)$ = (0.1,0.1).   
 The point, which  shows 
   the intersection of the solid curve and the dotted curve,  
  is given by   ($V_2/t$,$V/t$)=(1.15,2.0). 
  We find, within the numerical accuracy, 
     that the boundary between the state I and II 
      (the solid curve) is  expressed as $V = V_2  + const.$.  
 Such a $V_2$-dependence  is understood from the coefficient of 
    $D_{2Q_0}$  in  eq. (\ref{MFH}), i.e.,    $(U/2 -2 V + 2V_2)$.       
   The quantity,  $V$, which is needed to create  4$\kf$-CDW,   
     is replaced by $V-V_2$ for $V_2 \not= 0$. 
  The dotted curve denotes  a critical value of $V_2/t$,  above which    
   the 2$\kf$-CDW appears.  
  The tangent of the dotted curve depends on $\delta V$.  
 In  the inset, a phase diagram is shown 
  for $U/t = 4$ and  $(\td, \delta V)$ = (0,0). 
 The dotted curve becomes straight due to $\delta V = 0$, 
   since  the boundary between the state I and the state III 
      are determined only by 
   $V_2/t$ in the absence of $\td$  and $\delta V$. 
  It turns out that the 2$\kf$-CDW is induced by the next-nearest-neighbor 
   interaction but not by the nearest-neighbor interaction.
 The second term of eq. (\ref{MFH}) shows that 
  the coefficients, $(U/2 - 2V_2)$, of $D_{Q_0}$  changes 
 the sign  at $V_2/t=1$ for $U/t = 4$  and 
 that  the  2$\kf$-CDW is induced for larger $V_2/t$  
  {\it e.g.},    $D_1 \not= 0$   for $V_2/t \gsim 1.3$.
 Further we comment on the condition for $C(\pi/2) = 0$, which occurs 
 on the thin dash-dotted curve in Figs. 4 and Fig. 11(b). 
 The inset of Fig. 11(b) displaying  $C(\pi/2) = 0$ at $V_2=U/4$, 
  is interpreted from the fact that   
  the  coefficient  of $D_{Q_0}$ in the second line of eq. (\ref{MFH})
 becomes zero  at $V_2=U/4$
 resulting in the zero value of 2$\kf$-CDW.  
 
Finally we discuss the experimental results  based on our result. 
 In the present calculation, the state, which exhibits the coexistence of  
  the  2$\kf$-SDW and  2$\kf$-CDW\cite{Kobayashi} 
    followed by   4$\kf$-SDW, has been obtained 
   for a model with the on site, nearest-neighbor 
      and next-nearest-neighbor interactions, 
        i.e., the  purely electronic interactions. 
  The experimental results of the coexistence of 
    2$\kf$-SDW and 2$\kf$-CDW in TMTSF-salts\cite{Pouget} 
      seems to be consistent with the calculation.
 However  the experiment indicates also the  4$\kf$-CDW, 
    which is absent in the present calculation. 
 Such a discrepancy between the experiment and 
   the theory may suggest 
 a role of the electron-phonon interaction, which gives rise to 
  the density wave  in addition to the purely  electronic  interaction. 
 Further, it will be an interesting problem to elucidate if 
   4$\kf$-SDW obtained in the present calculation is observed  
      in the coexistent state of the Bechgaard  salts.

%\section*{ Acknowledgments }
\vspace{1cm}
%------------Acknowledgment--------
{\bf Acknowledgment}  \\
 This work was partially supported by  a Grant-in-Aid 
 for Scientific  Research  from the Ministry of Education, 
Science, Sports and Culture (Grant No.09640429).

%\newpage

%-------------  References              ----

\end{document}